\documentclass[a4paper,11pt]{article}
\usepackage{pos}
\usepackage[utf8]{inputenc}
\usepackage{amsmath}
\usepackage{amsfonts}
\usepackage{subcaption}

\newcommand{\Tr}[1]{\text{Tr}\left[#1\right]}

\title{Tensor renormalization group study of 3D principal chiral model}
\author[a,b]{Shinichiro Akiyama}
\author[c]{Raghav G.~Jha}
\author*[d]{Judah Unmuth-Yockey}
\affiliation[a]{Center for Computational Sciences, University of Tsukuba, Ibaraki, 305-8577, Japan}
\affiliation[b]{Endowed Chair for Quantum Software, University of Tokyo, Tokyo, 113-0033, Japan}
\affiliation[c]{Thomas Jefferson National Accelerator Facility, Newport News, VA 23606, USA}
\affiliation[d]{Department of Theoretical Physics, Fermi National Accelerator Laboratory, Batavia IL, USA}

\abstract{
We study the three-dimensional $SU(2)$ principal chiral model (PCM) using different tensor renormalization group methods based on the triad and anisotropic decomposition of the tensor. The tensor network representation is formulated based on the character expansion of the Boltzmann weight. We compare the average action obtained using these two tensor network algorithms and confirm that the resulting critical coupling and exponent are comparable with the recent estimations from the Monte Carlo methods.
}

\FullConference{The 40th International Symposium on Lattice Field Theory (Lattice 2023)\\
July 31st - August 4th, 2023\\
Fermi National Accelerator Laboratory\\}

\notes{
    \note{UTCCS-P-151}
}

\begin{document}
\maketitle

\section{Introduction}

Tensor network methods possess several enjoyable properties:  First, they are very efficient for large volumes with translational symmetries.  
In those cases, some of the methods allow us to handle exponentially large lattice sizes just in polynomial time.
Second, tensor networks are sign-problem-free in the sense that there is no probabilistic sampling, so negative or imaginary weights pose no challenge. These nice features are not without some difficulties.  One of the biggest hurdles for the tensorial approach is to extend it to higher dimensions.  This is attributed to the poor scaling of the algorithms with space-time dimension.  In two Euclidean dimensions tensor network methods are well understood, fast, and state-of-the-art, however, in three and four dimensions they are not understood well. 

There are several algorithms that are tailored for higher dimensions~\cite{PhysRevB.97.045111,PhysRevB.86.045139,verstraete2004renormalization,kadoh2019renormalization,PhysRevA.74.022320,PhysRevB.90.125154,PhysRevB.102.054432}. In this study we focus on two of these in particular: triad TRG (tTRG)~\cite{kadoh2019renormalization}, and anisotropic TRG (ATRG)~\cite{PhysRevB.102.054432}, and compare them on a non-trivial $SU(2)$ principal chiral model (PCM).  This comparison will help demonstrate the benefits of the different algorithms along with the drawbacks inherent to the approaches.

For the comparison to be useful, we consider a model with some degree of generality. That is, it should share some common features with other physically interesting models.  In addition, it should be checkable in various ways, either through alternative numerical methods or through analytical means.  A model that stands out is the three-dimensional $SU(2)$ PCM.  This model is equivalent to the $O(4)$ nonlinear sigma model.  This model may be an effective theory for two-flavor, large temperature quantum chromodynamics~\cite{PhysRevD.20.2610,PhysRevD.29.338,PhysRevD.55.362,PhysRevD.61.054503}, as well as an effective theory for large temperature, strong coupling 4D $SU(2)$ gauge theory.

\section{Tensor formulation of $SU(2)$ principal chiral model}

The action for the $SU(2)$ PCM in the continuum can be transcribed onto the lattice as,
\begin{align}
	S_{\rm cont}
	=
	\frac{\beta}{2}
	\int d^{3}x
	\Tr{
		\sum_{\nu=1}^{3}\partial_{\nu}U(x)^{\dag}\partial_{\nu}U(x)
	} \rightarrow S_{\rm lat} = -\frac{\beta}{2}
	\sum_{n,\nu}
	\Re\left\{
	\Tr{
	U(n)U(n+\hat{\nu})^{\dag}
	}
	\right\}
\end{align}
where $U(x)$ are $SU(2)$ matrices, $n$ are the sites of the lattice and $\beta$ is the coupling.  
The lattice action can then be used to define a statistical mechanics partition function in the canonical ensemble,
\begin{align}
\label{eq:path_int}
	Z= \int
	\left(
	\prod_{n}d U(n)
	\right)
	e^{-S_{\rm lat}}
\end{align}
where $dU(n)$ is the Haar integration measure over $SU(2)$. Due to the continuous nature of the $U$ matrices, an immediate transcription of the partition function into a tensor network is not possible.  Instead, we can use the compact nature of the group to expand the nearest-neighbor Boltzmann weight,
\begin{align}
      e^{-\frac{\beta}{2} \Re{\Tr{U(x) U^{\dagger}(x + \hat{\nu})}}}
    =
    \sum_{r = 0}^{\infty} F_{r}(\beta) \chi^{r}(U(x) U^{\dagger}(x + \hat{\nu})),
\end{align}
where $r$ label the half-integer irreducible representations (irreps.) of the $SU(2)$ group, $F_{r}(\beta)$ are the expansion coefficients given in terms of modified Bessel functions, and $\chi^{r}$ are the characters of the group. This expansion introduces discrete, half-integer degrees of freedom associated with the interaction surfaces of the lattice---in this case the links.
The characters can be used to isolate the group matrices using,
\begin{align}
    \chi^{r}(U(x) U(x+\hat{\nu})) = \sum_{m,n} D_{m n}^{r}(U(x)) {D^{r}_{n,m}}^{\dagger}(U(x+\hat{\nu})).
\end{align}
This factorization allows for exact integration over the original, matrix degrees of freedom, leaving only the discrete degrees of freedom behind.  This integration creates constraints between the different irreps. in the form of Clebsch-Gordon coefficients. After integrating out all the original matrix variables a local tensor can be constructed identically at every site on the lattice having the form,
\begin{align}
    \nonumber
	&
	T_{
	(r_{1}m_{1}n_{1})
	(r_{2}m_{2}n_{2})
	(r_{3}m_{3}n_{3})
	(r_{4}m_{4}n_{4})
	(r_{5}m_{5}n_{5})
	(r_{6}m_{6}n_{6})
	}
	=
	\sqrt{\prod_{p=1}^{6}F_{r_{p}}(\beta)}
	\nonumber\\
	&\times
	\sum_{R_{12}=|r_{1}-r_{2}|}^{r_{1}+r_{2}}
	\sum_{R_{123}=|R_{12}-r_{3}|}^{R_{12}+r_{3}}
	\sum_{R_{56}=|r_{5}-r_{6}|}^{r_{5}+r_{6}}
	\sum_{M_{12},N_{12}}
	\sum_{M_{123},N_{123}}
	\sum_{M_{56},N_{56}}
        \frac{1}{2R_{123}+1}
	\nonumber\\
	&\times
	C^{R_{12}M_{12}}_{r_{1}m_{1}r_{2}m_{2}}
	C^{R_{12}N_{12}}_{r_{1}n_{1}r_{2}n_{2}}
	C^{R_{123}M_{123}}_{R_{12}M_{12}r_{3}m_{3}}
	C^{R_{123}N_{123}}_{R_{12}N_{12}r_{3}n_{3}}
	C^{R_{123}M_{123}}_{r_{4}m_{4}R_{56}M_{56}}
	C^{R_{123}N_{123}}_{r_{4}n_{4}R_{56}N_{56}}
	C^{R_{56}M_{56}}_{r_{5}m_{5}r_{6}m_{6}}
	C^{R_{56}N_{56}}_{r_{5}n_{5}r_{6}n_{6}}
        .
\end{align}
\begin{figure}
    \centering
    \includegraphics[width=0.75\textwidth]{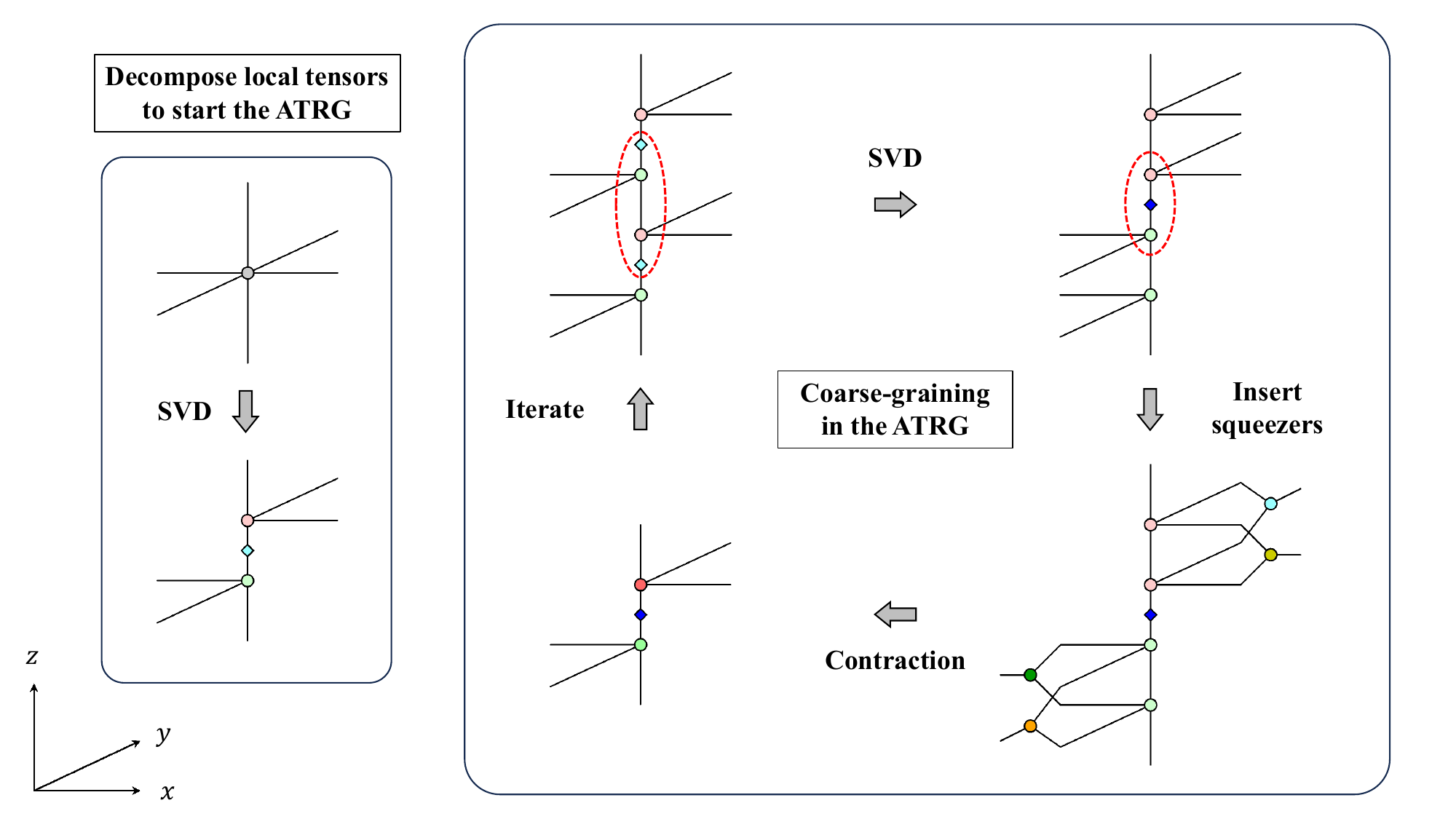}
    \caption{The decomposition of the initial tensor (left) and  schematic overview of the ATRG algorithm (right) as proposed in Ref.~\cite{PhysRevB.102.054432}.}
    \label{fig:atrg-update}
\end{figure}
The ATRG uses a decomposition of the initial tensor given by,
\begin{align}
  &T_{
	(r_{1}m_{1}n_{1})
	(r_{2}m_{2}n_{2})
	(r_{3}m_{3}n_{3})
	(r_{4}m_{4}n_{4})
	(r_{5}m_{5}n_{5})
	(r_{6}m_{6}n_{6})
	}
	\nonumber\\
	&\approx
	\sum_{\gamma}
	U_{
	(r_{1}m_{1}n_{1})
	(r_{2}m_{2}n_{2})
	(r_{3}m_{3}n_{3})
	\gamma
	}
	\sigma_{\gamma}
	V^{*}_{
	(r_{4}m_{4}n_{4})
	(r_{5}m_{5}n_{5})
	(r_{6}m_{6}n_{6})
	\gamma
	}
\end{align}
which can be seen in the left-hand side of Fig.~\ref{fig:atrg-update}.
This initial singular value decomposition (SVD) of the tensor is used in the main updating step of the algorithm, as can be seen in the right-hand side of Fig.~\ref{fig:atrg-update}.  The top and bottom of adjacent tensors are swapped via SVD in the algorithm, which allows for an algorithm that can be iterated.

\begin{figure}
    \centering
    \begin{subfigure}[t]{0.4\textwidth}
        \centering
    \includegraphics[width=\textwidth,trim={1cm 10cm 1cm 9cm},clip]{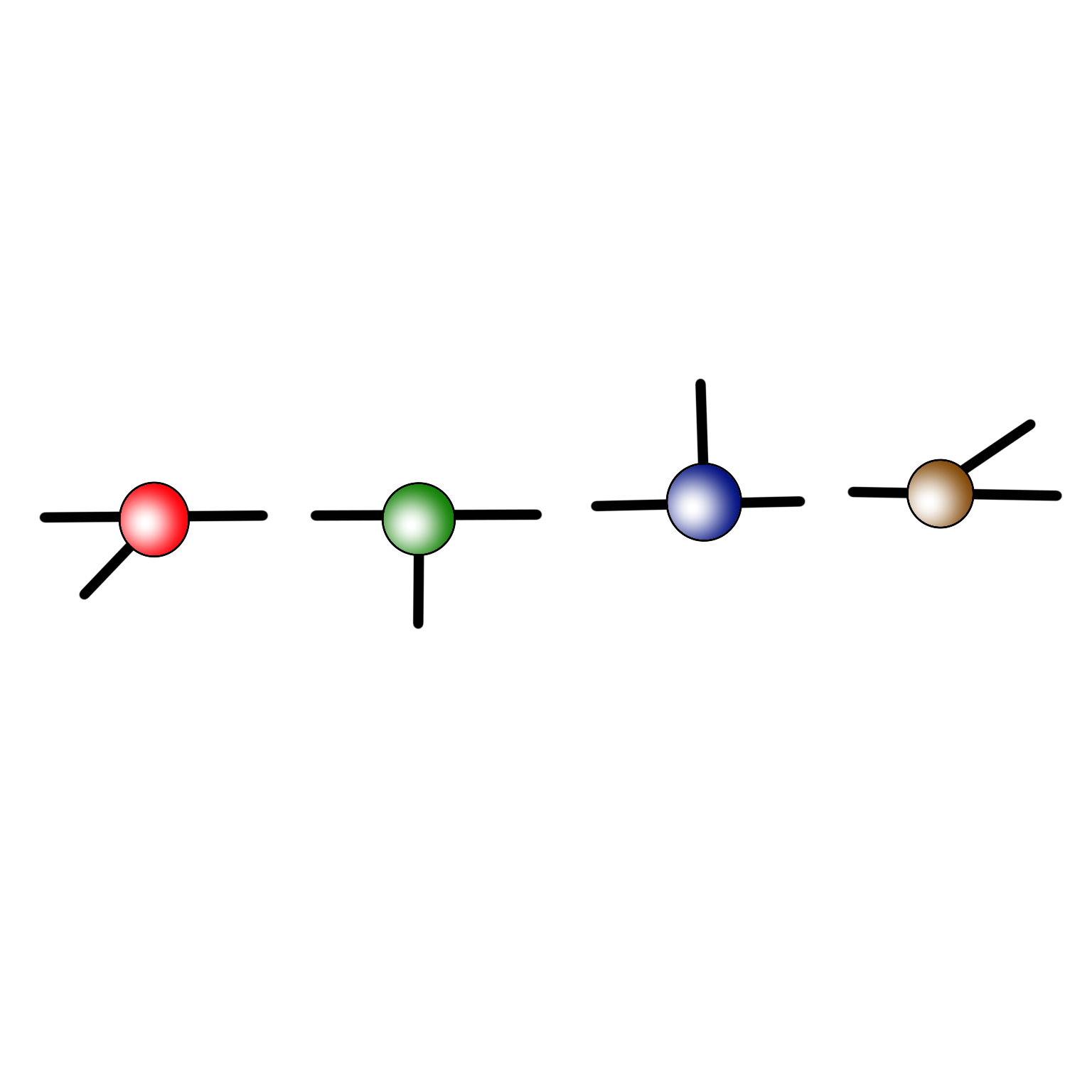}
    \caption{An illustration of the decomposition of the initial tensors using triads.}
    \label{fig:triad-decomp}
    \end{subfigure}
    \hfill
    \begin{subfigure}[t]{0.4\textwidth}
        \centering
    \includegraphics[width=\textwidth,trim={4cm 4cm 4cm 3.5cm},clip]{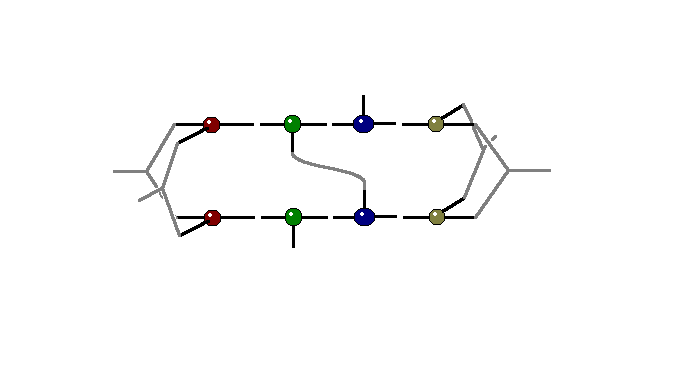}
    \caption{The higher-order tensor renormalization group~\cite{PhysRevB.86.045139} update applied to the tTRG.}
    \label{fig:triad-update}
    \end{subfigure}
    \caption{Schematic picture of tTRG algorithm~\cite{kadoh2019renormalization}.}
\end{figure}

The tTRG uses a decomposition of the initial tensor into four smaller tensors.  This decomposition can be approximate.
\begin{equation}
    T_{i j k l m n}=\sum_{a, b, c} A_{i k a} B_{a m b} C_{b n c} D_{c l j}
\end{equation}
Figure~\ref{fig:triad-decomp} gives an illustration of the four tensors and their connectivity.
In the case of the $SU(2)$ PCM, the four tensors are given by
\begin{align}
& \hspace{-0.7cm} A_{\left(r_{1}, m_{1}, n_{1}\right) \left(r_{2}, m_{2}, n_{2}\right) (R, M, N)}=\sqrt{F_{r_{1}}(\beta) F_{r_{2}}(\beta)} C_{r_{1} m_{1} r_{2} m_{2}}^{R M} C_{r_{1} n_{1} r_{2} n_{2}}^{R N} \\ 
& \hspace{-0.7cm} B_{(R, M, N) \left(r_{3}, m_{3}, n_{3}\right)\left(R^{\prime}, M^{\prime}, N^{\prime}\right)}=\frac{1}{\sqrt{d_{R^{\prime}}}} \sqrt{F_{r_{3}}(\beta)} C_{R M r_{3} m_{3}}^{R^{\prime} M^{\prime}} C_{R N r_{3} n_{3}}^{R^{\prime} N^{\prime}} \\ 
& \hspace{-0.7cm}  C_{\left(R^{\prime}, M^{\prime}, N^{\prime}\right)\left(r_{4}, m_{4}, n_{4}\right) \left(R^{\prime \prime}, M^{\prime \prime}, N^{\prime \prime}\right)}=\frac{1}{\sqrt{d_{R^{\prime}}}} \sqrt{F_{r_{4}}(\beta)} C_{R^{\prime \prime} M^{\prime \prime} r_{4} m_{4}}^{R^{\prime} M^{\prime}} C_{R^{\prime \prime} N^{\prime \prime} r_{4} n_{4}}^{R^{\prime} N^{\prime}} \\
& \hspace{-0.7cm}  D_{\left(R^{\prime \prime}, M^{\prime \prime}, N^{\prime \prime}\right) \left(r_{5}, m_{5}, n_{5}\right) \left(r_{6}, m_{6}, n_{6}\right)}=\sqrt{F_{r_{5}}(\beta) F_{r_{6}}(\beta)} C_{r_{5} m_{5} r_{6} m_{6}}^{R^{\prime \prime} M^{\prime \prime}} C_{r_{5} n_{5} r_{6} n_{6}}^{R^{\prime \prime} N^{\prime \prime}}
,
\end{align}
where the $C$s are Clebsch-Gordon coefficients, and $d_r$ is the dimension of the irrep.  The algorithm updates the tensors iteratively using the HOTRG algorithm, and the update step can be seen in Fig.~\ref{fig:triad-update}.

\section{Observables \& Results}

To compare the tTRG and the ATRG we compare several observables.  A simple quantity to extract from most tensor network algorithms is the free energy density given by (up to a sign),
\begin{align}
    F \equiv \frac{1}{V} \log(Z),
\end{align}
where $Z$ is the partition function, and $V$ is the system volume.    The average action and its susceptibility are given by,
\begin{align}
    \langle s \rangle = \frac{\langle S \rangle}{V} \equiv -\frac{\partial}{\partial \beta} F, \quad 
    \chi_{s} = V (\langle s^{2} \rangle - \langle s\rangle^{2}) \equiv \frac{\partial^{2}}{\partial \beta^{2}} F.
\end{align}
Another useful observable first proposed in Ref.~\cite{PhysRevB.80.155131} which effectively counts the ground state degeneracy is defined as: 
\begin{align}
    X \equiv \frac{\left( T_{aabbcc} \right)^{2}}{T_{abccdd} T_{baeeff}},
    \label{eq:X_def} 
\end{align}
where the tensor indices are ordered \emph{left, right, front, back, top, bottom}, and repeated indices are summed.  This quantity can be computed in each direction during the update step in the tensor algorithm.  
\begin{figure}
\centering
\begin{subfigure}[t]{0.45\textwidth}
    \includegraphics[width=\textwidth]{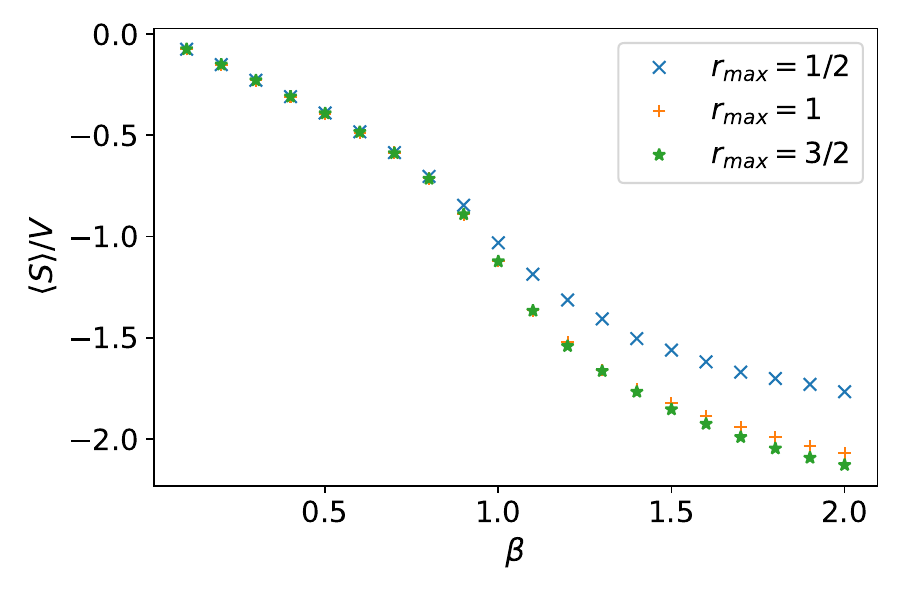}
\caption{ATRG, $D = 40$, $V = 1024^{3}$.}
\label{fig:atrg-action}
\end{subfigure}
\hfill
\begin{subfigure}[t]{0.45\textwidth}
    \includegraphics[width=\textwidth]{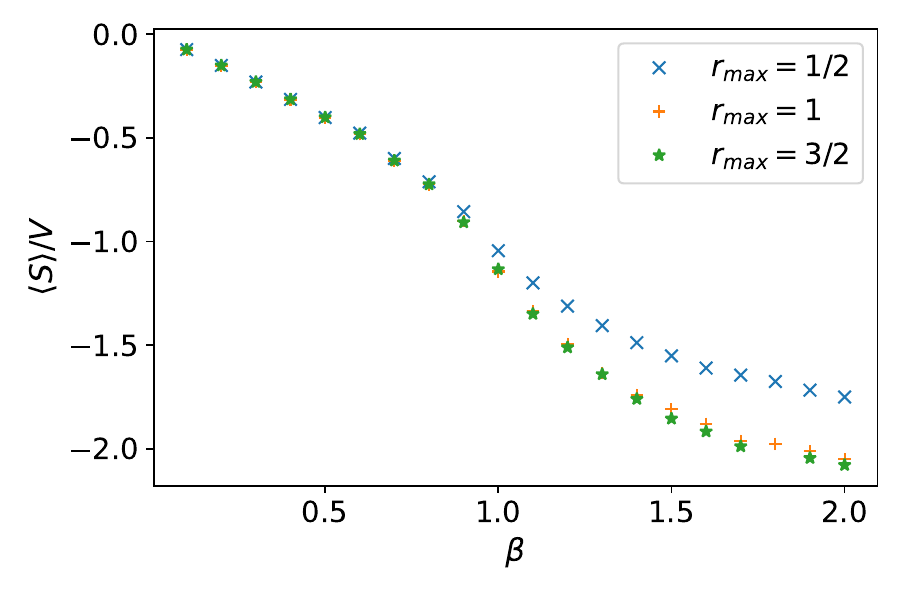}
\caption{Triad, $D = 40$, $V = 1024^{3}$.}
\label{fig:triad-action}
\end{subfigure} 
\caption{Average action varying $r_{\rm max}$.}
\end{figure} 
The results for the average action using ATRG and tTRG are shown in Figs.~\ref{fig:atrg-action} and~\ref{fig:triad-action} for three different truncations across a range of $\beta$ values. These calculations were done at $V = 1024^{3}$ with $D=40$.  
\begin{figure}
\centering
\begin{subfigure}[t]{0.49\textwidth}
    \centering
    \includegraphics[width=\textwidth]{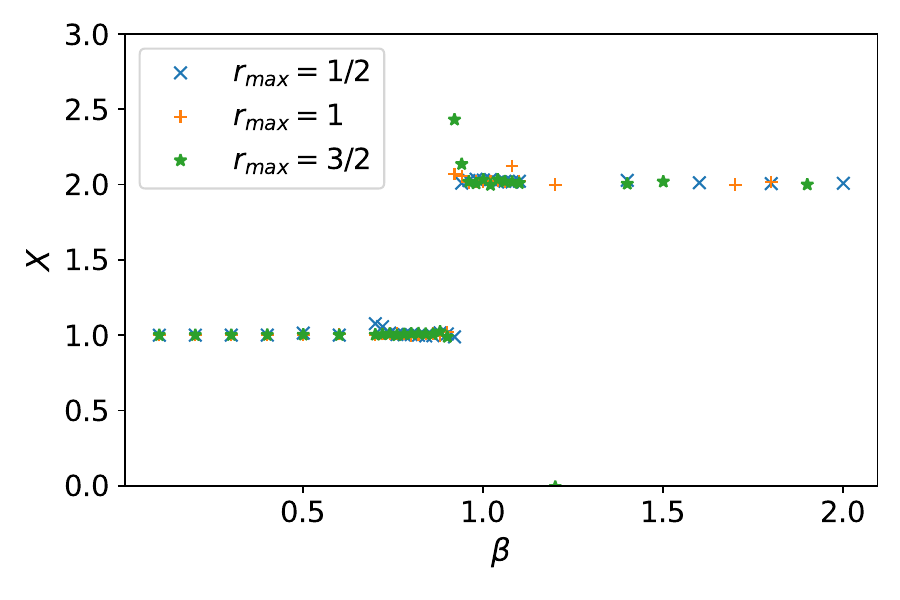}
\caption{Triad, $X$ varying $r_{\text{max}}$ and $\beta_{c}$, $D = 40$, $V = 1024^{3}$. \\
  $r_{\text{max}} = 1/2$: $\beta_{c} = 0.935(5)$, 
  $r_{\text{max}} = 1$: $\beta_{c} = 0.915(5)$, 
  $r_{\text{max}} = 3/2$: $\beta_{c} = 0.915(5)$}
  \label{fig:x-triad}
\end{subfigure}
\hfill
\begin{subfigure}[t]{0.49\textwidth}
    \centering
\includegraphics[width=\textwidth]{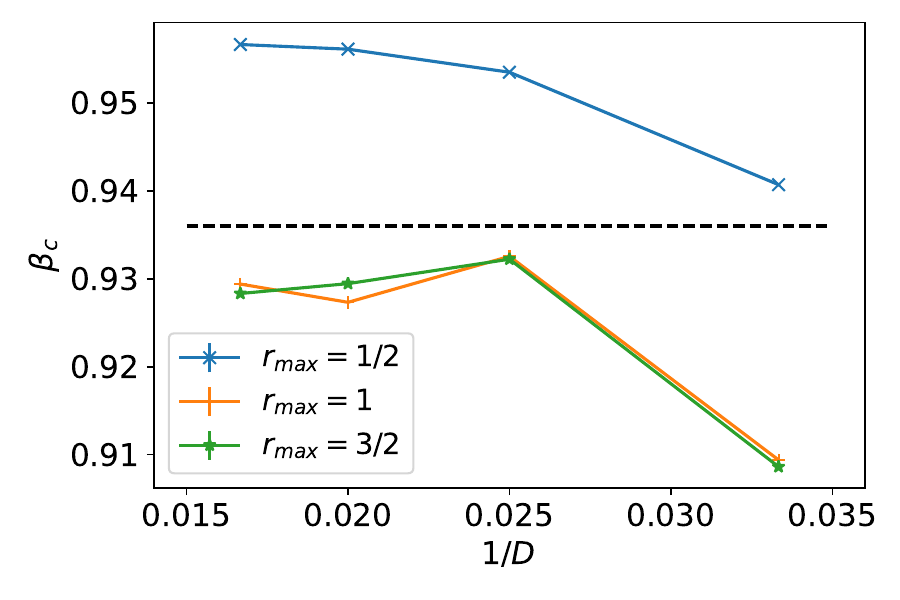}
\caption{$\beta_{c}$ from ATRG using $X$, $V = 1024^{3}$. The recent MC estimation is $\beta_{c} = 0.9360(1)$~\cite{L_pez_Contreras_2022} shown by the dashed line.}
\label{fig:x-atrg}
\end{subfigure}
\caption{Estimation of the critical coupling using $X$ defined in \eqref{eq:X_def}}. 
\end{figure}

\begin{figure}
\centering
\begin{subfigure}[t]{0.3\textwidth}
    \centering
    \includegraphics[width=\textwidth]{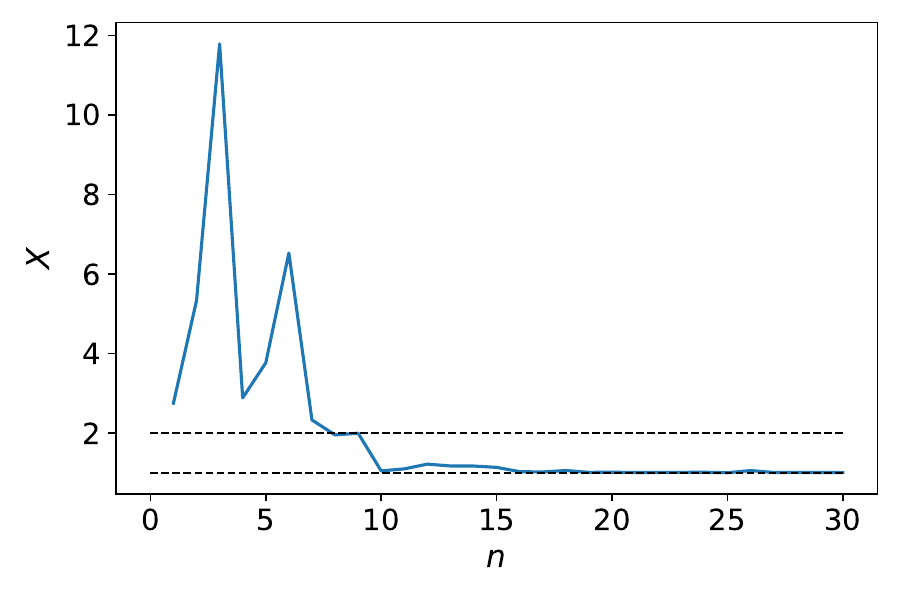}
    \caption{$\beta = 0.8$, $r_{\rm max} = 1/2$}    
\end{subfigure}
\begin{subfigure}[t]{0.3\textwidth}
    \centering
    \includegraphics[width=\textwidth]{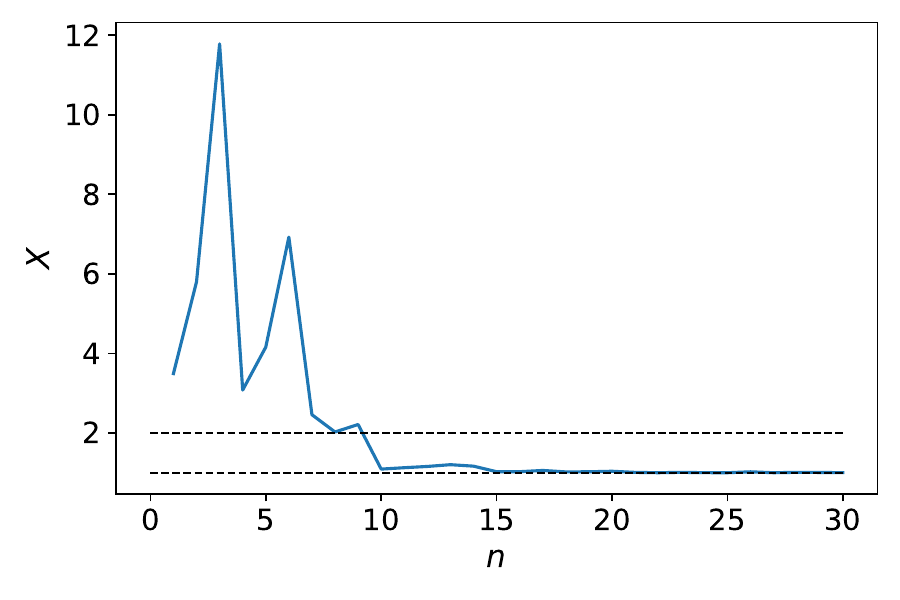}
\caption{$\beta = 0.8$, $r_{\rm max} = 1$}    
\end{subfigure}
\begin{subfigure}[t]{0.3\textwidth}
    \centering
    \includegraphics[width=\textwidth]{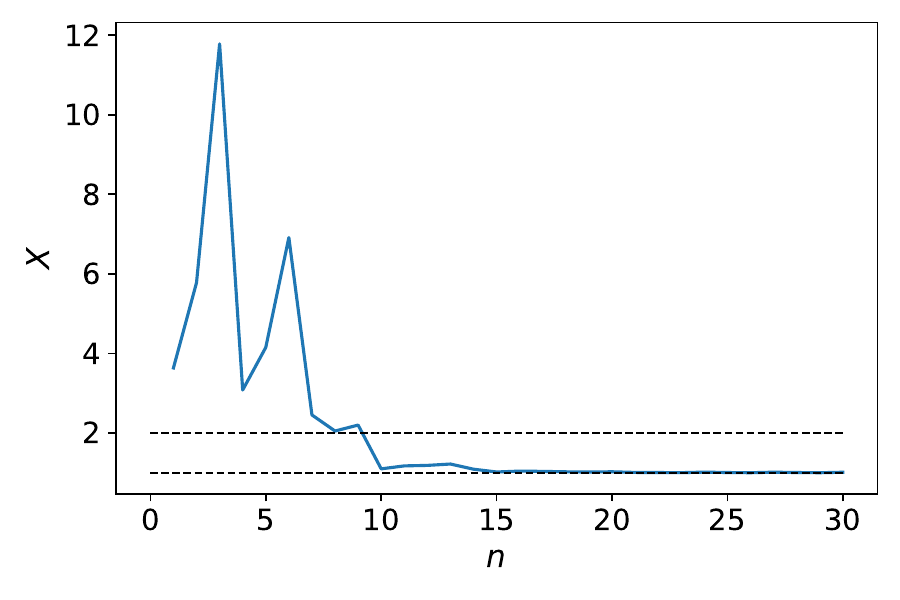}   
    \caption{$\beta = 0.8$, $r_{\rm max} = 3/2$}    
\end{subfigure}
\begin{subfigure}[t]{0.3\textwidth}
    \centering
    \includegraphics[width=\textwidth]{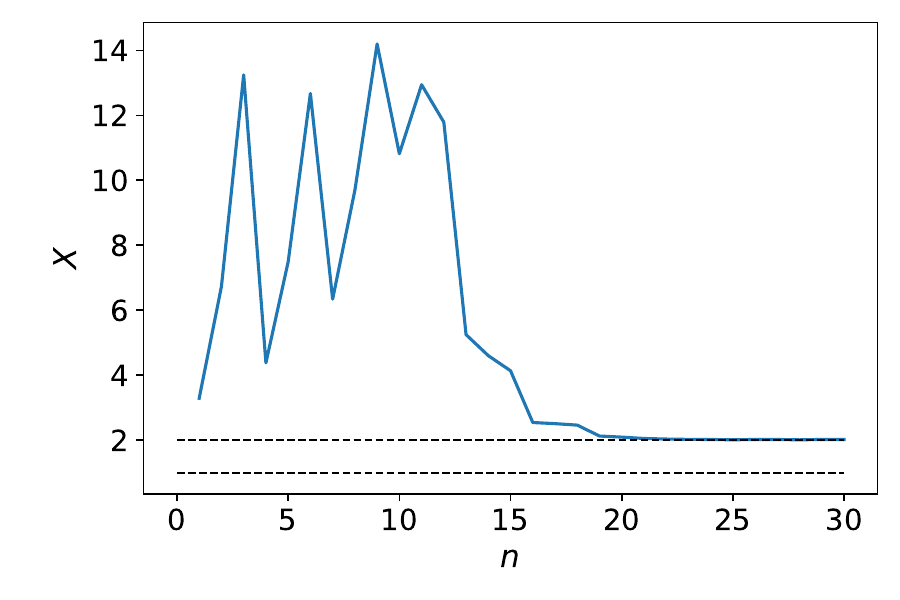}        
    \caption{$\beta = 1.1$, $r_{\rm max} = 1/2$}    
\end{subfigure}
\begin{subfigure}[t]{0.3\textwidth}
    \centering
    \includegraphics[width=\textwidth]{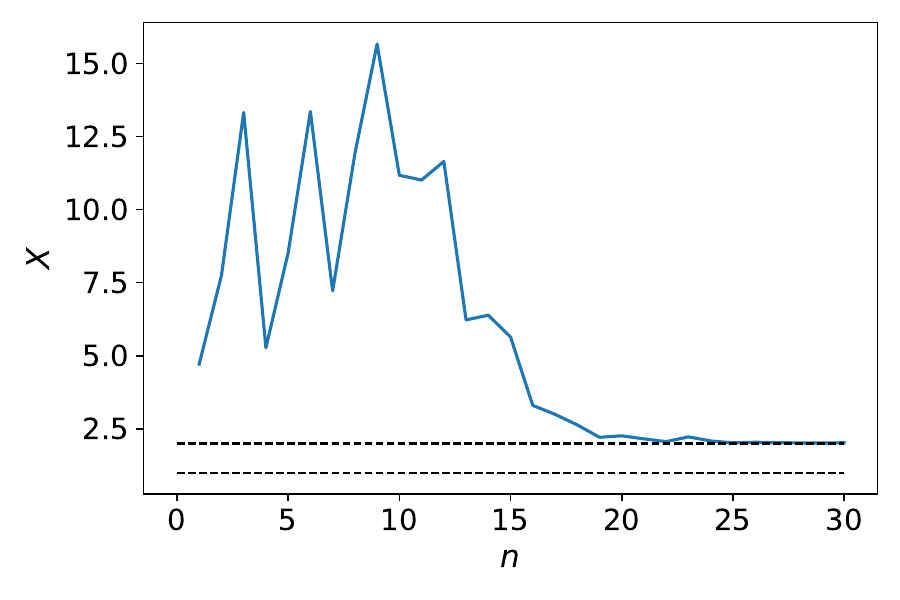}        
    \caption{$\beta = 1.1$, $r_{\rm max} = 1$}    
\end{subfigure}
\begin{subfigure}[t]{0.3\textwidth}
    \centering
    \includegraphics[width=\textwidth]{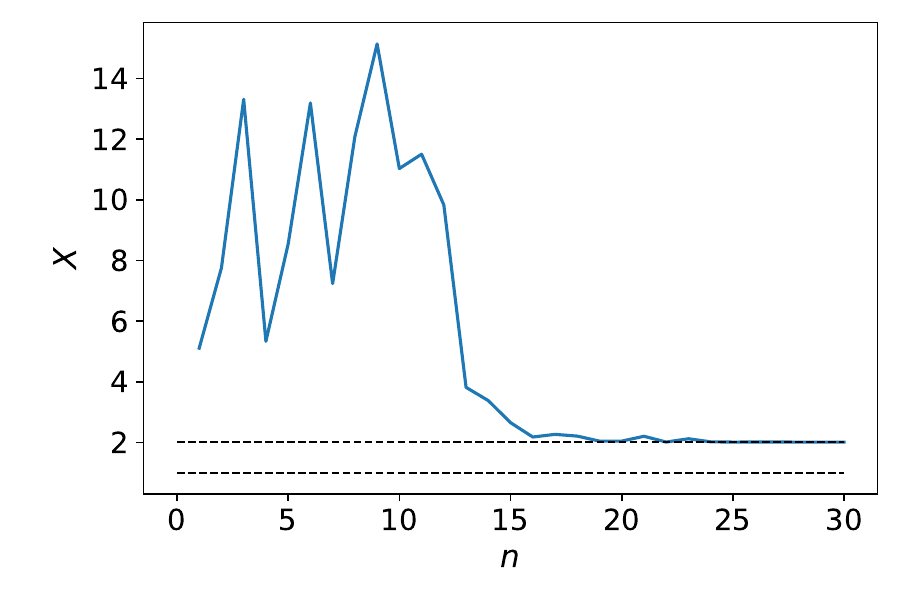}        
    \caption{$\beta = 1.1$, $r_{\rm max} = 3/2$}    
\end{subfigure}
\caption{The convergence of $X$ with the number of coarse-graining step $n$, for two different $\beta$ values, and three different truncations.}
\label{fig:triad-x-iteration}
\end{figure}

To locate the phase transition in tensor computations, one usually has to compute the derivative of some observable. However, using $X$, we can locate transitions without explicitly computing any derivative. This is computationally easier. We computed $\beta_{c}$ using the discontinuous behavior of the observable, $X$. The result is shown in Fig.~\ref{fig:x-triad} using the tTRG for three different values of the irrep. cut-off.  We find the critical coupling for $r_{\text{max}} = 1/2$ differs from the next two highest cut-offs.  Figure~\ref{fig:x-atrg} shows similar results obtained using the ATRG.  In this case, the results are shown as a function of the inverse tensor network bond dimension for the same three values of truncation.  The convergence of $X$ as a function of the iteration step of the tTRG can be seen in Fig.~\ref{fig:triad-x-iteration}. For sufficiently small $\beta$, $X$ converges to 1, while for large $\beta$ it converges to 2.

Using the data for the average action, along with the results from calculating $X$, it is possible to extract a critical exponent $\alpha$ using the fitting ansatz
\begin{align}
        \langle S \rangle / V = A &+ B|\beta -\beta_{c}|
        + C|\beta -\beta_{c}|^{1-\alpha}.
        \label{eq:ansatz_fit}
\end{align}
\begin{figure}
\centering
\begin{subfigure}[t]{0.49\textwidth}
    \centering
    \includegraphics[width=\textwidth]{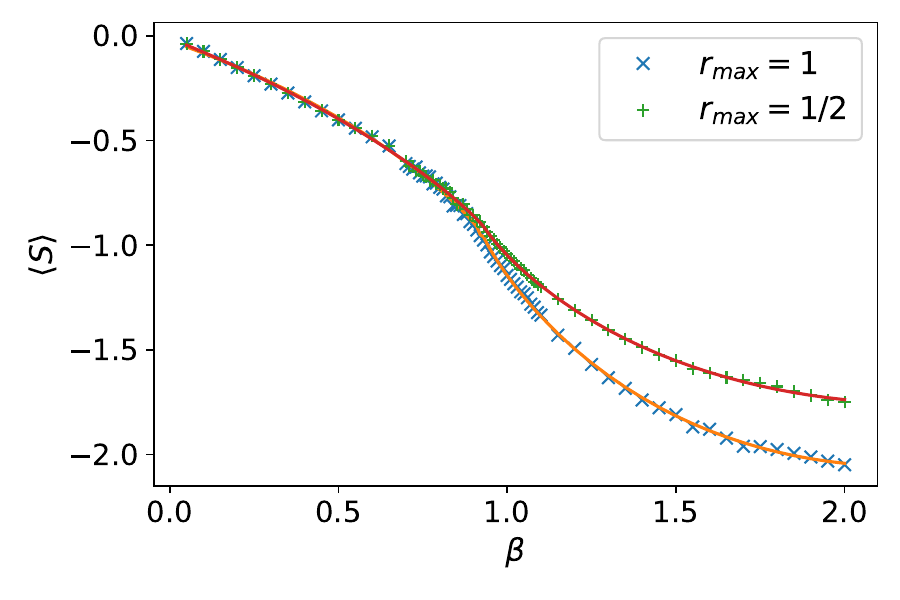}        
\caption{Triad low-high temp fit, $D = 40$, $V = 1024^{3}$. \\
      $r_{\rm max}=1/2$: $\beta_{c} = 0.933(1)$, $r_{\rm max} = 1$: $\beta_{c} = 0.9232(1)$,
$r_{\rm max} = 3/2$: $\beta_{c} = 0.92295(4)$}
\label{fig:triad-fit}
\end{subfigure}
\hfill
\begin{subfigure}[t]{0.49\textwidth}
    \centering
\includegraphics[width=\textwidth]{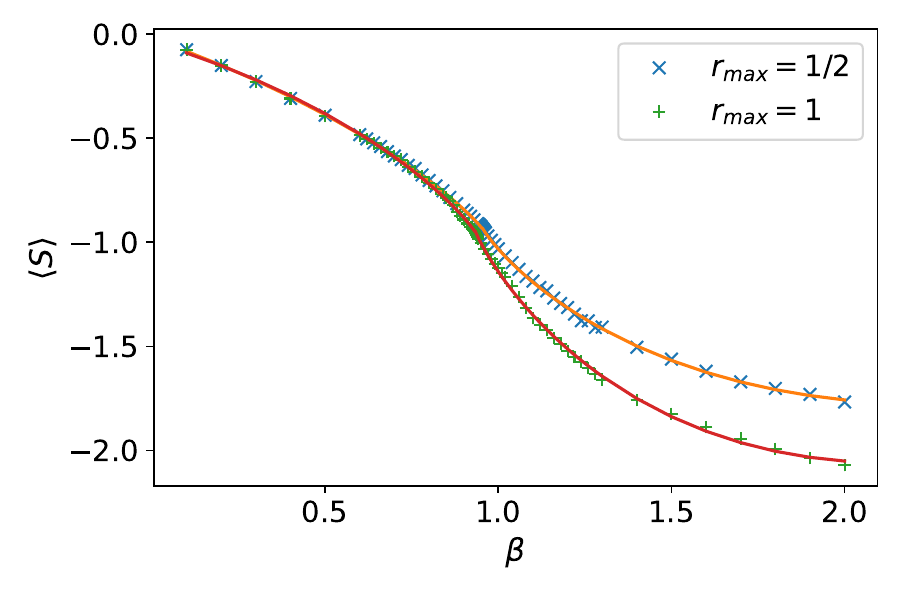}
\caption{ATRG low-high temp fit, $D=40$, $V = 1024^{3}$. \\
$r_{\text{max}} = 1/2$: $\alpha = -0.21259(1)$,
$r_{\text{max}} = 1$: $\alpha = -0.2151(1)$,
$r_{\text{max}} = 3/2$: $\alpha = -0.22676(4)$}
\label{fig:atrg-fit}
\end{subfigure} 
\caption{Fitting the ansatz of \eqref{eq:ansatz_fit} for the average action to the numerical results obtained from tTRG and ATRG at fixed bond dimension and lattice volume.}
\end{figure}
Figures~\ref{fig:triad-fit} and ~\ref{fig:atrg-fit} show results from fitting the above ansatz to the data.  In the case of the tTRG, we first fix the value of $\alpha$ using the literature, $\alpha = -0.247(6)$~\cite{Francesco_Parisen_Toldin_2003}, then fit for $\beta_{c}$.  The results in Fig.~\ref{fig:triad-fit} are to be compared with the value $\beta_{c} = 0.9360(1)$~\cite{L_pez_Contreras_2022}.  For the ATRG we use the results from $X$ for the value of $\beta_{c}$ and fit to $\alpha$.  In comparison to the calculation of $X$, again we find both with the tTRG and the ATRG that the $r_{\text{max}}=1/2$ case differs from the higher cut-off cases.

\section{Conclusion}

In our study, we find that both tTRG and the ATRG can be useful.  The ATRG is more accurate at a fixed bond dimension, however, it is also computationally more expensive.  Both methods seem to locate the phase transition reasonably well. In addition, our results show for the first time that the identification of the phase transition using $X$ which had only been used for models with discrete groups can be extended to models with continuous symmetry. The results from $X$ seem to indicate the transition is associated with a $\mathbb{Z}_{2}$ symmetry.  To understand this better it would be interesting to calculate $\langle \Tr U \rangle$ and its susceptibility.  In addition, studying the $SU(3)$ PCM would increase the complexity of the model allowing a study of the group $SU(3)$ and its character expansion, which is crucial for later work on quantum chromodynamics. We leave these questions for future explorations. 

\begin{acknowledgments}
SA acknowledges the support from the Endowed Project for Quantum Software Research and Education, the University of Tokyo (\url{https://qsw.phys.s.u-tokyo.ac.jp/}) 
and JSPS KAKENHI Grant Number JP23K13096. A part of ATRG calculations for the present work was carried out with ohtaka provided by the Institute for Solid State Physics, the University of Tokyo.  RGJ is supported by the U.S. Department of Energy, Office of Science, National Quantum Information Science Research Centers, Co-design Center for Quantum Advantage (C2QA) under contract number DE-SC0012704 and by the U.S. Department of Energy, Office of Science, Office of Nuclear Physics under contract number DE-AC05-06OR23177. This work is supported by the Department of Energy through the Fermilab Theory QuantiSED program in the area of “Intersections of QIS and Theoretical Particle Physics.  This manuscript has been authored by Fermi Research Alliance, LLC under Contract No. DE-AC02-07CH11359 with the U.S. Department of Energy, Office of Science, Office of High Energy Physics.
The authors would like to thank the Institute for Nuclear Theory at the University of Washington for its kind hospitality and stimulating research environment. Discussions during and after the INT WORKSHOP INT-21R-1C on "Tensor Networks in Many Body and Quantum Field Theory" were useful to initiate this work.
\end{acknowledgments}


\bibliographystyle{utphys}
\raggedright
\providecommand{\href}[2]{#2}\begingroup\raggedright
\endgroup
\end{document}